# Optimal designs for incomplete stepped wedge trials

Richard Hooper[1], Alan Girling[2]

[1] Wolfson Institute of Population Health, Queen Mary University of London, London, UK

[2] Institute of Applied Health Research, University of Birmingham, Birmingham, UK

## Abstract

*Background*

Stepped wedge trials are longitudinal randomised evaluations, usually cluster-randomised, in which the experimental intervention is introduced in a staggered fashion. Incomplete stepped wedge designs focus the effort of data collection on particular periods in particular sequences.

*Methods*

We suppose there is a cost for every period in every cluster where we collect data, and that there are a fixed number of individuals, $m$, with data available in each period in each cluster. If we are willing to pay the cost of data collection in that cluster-period then we collect the data on all $m$ individuals, and if we are not willing to pay the cost then we collect no data in that cluster-period. We consider the problem of designing a trial to minimise the total number of cluster-periods of data collection needed to achieve given precision for the treatment effect estimator, or equivalently, to maximise precision for a given number of cluster-periods of data collection.

*Results*

We present the solution for two-period trials, which depends on the correlation, $\rho$, between two cluster-period means from the same cluster in different periods. There are two distinct forms for the solution, depending on whether $\rho < \sqrt{3}/2$. We also present a conjecture on the form of the solution for multi-period trials, informed by results from a greedy search of the design space.

*Conclusions*

A real-life stepped wedge design problem will involve trading off the costs of various design elements subject also to constraints on the scale of data collection. Nevertheless, the solutions to the problem considered here add significantly to our understanding of the optimal design of incomplete stepped wedge trials.

***Keywords:*** clinical trials, cluster randomised, incomplete design, longitudinal, minimum variance, stepped wedge, trial design

## Background

Stepped wedge trials are longitudinal randomised evaluations, usually cluster-randomised, in which the experimental intervention is introduced in a staggered timetable.[1] Rather than being randomised to the intervention or control conditions, clusters of individuals are randomised to one of a number of "sequences" in which they cross over unidirectionally at some point from the routine care (control) condition to the intervention condition.[2] In the most general framework, stepped wedge designs may also include sequences where clusters remain in the control condition or intervention condition for the duration of the trial – in fact, adding these sequences can improve the statistical efficiency of the design.[3,4]

Stepped wedge trials come in a variety of forms,[5,6] but the simplest kind to conceptualise involves taking a series of repeated, cross-sectional, independent samples from each cluster. The repeated cross-sections define discrete "periods" of time. Research on optimal design for repeated cross-sectional stepped wedge trials has mostly focused on "complete" designs—designs under which data are collected from each available cluster in each available period.[7] The typical motivation in the complete case is to imagine there is a cost associated with each cluster recruited to the trial, but no additional cost associated with the *amount* of individual-level data collected within that cluster (so that one might as well collect all of it). We know in this case how to design a complete, repeated cross-sectional stepped wedge trial that minimises the number of clusters needed to achieve given precision for the treatment effect estimate (at least under an exchangeable model for the between-period intra-cluster correlation).[4]

Incomplete stepped wedge designs are designs that restrict data collection to particular periods in particular sequences. A "sequence" in an incomplete design is not simply a sequence of treatment conditions: it also specifies which periods include data collection. Incomplete designs can be used, for example, to formalise the notion of a "transition period" (with no data collection) between intervals of data collection under the control condition and under the intervention condition.[1] They are also of interest if we want to improve efficiency of the trial by adopting a more targeted approach to data collection.[8] Suppose there is a cost for every period in every cluster where we collect data. Suppose, for simplicity, that there are a fixed number of individuals, $m$, with data available in each period in each cluster. If we are willing to pay the cost of data collection in that cluster-period then we collect the data on all $m$ individuals, and if we are not willing to pay the cost then we collect no data in that cluster-period. In this scenario we are interested in the following optimisation problem: how to design a trial to minimise the number of cluster-periods of data collection needed to achieve given precision (assuming there are no other constraints on the number of clusters we can include). The optimal design in this case may be an incomplete design.

In this article we derive solutions to this optimisation problem for the two-period case, and offer a conjecture on the form of solutions for designs with more than two periods, under an exchangeable model for the between-period intra-cluster correlation. But first we offer a motivating example.

### *Motivating example*

Consider a trial evaluating the effects of a free school breakfast programme in schools on pupils' cognitive performance and diet. We imagine a trial that shares some characteristics with the evaluation of the Welsh Assembly Government's Primary School Free Breakfast Initiative, conducted over the two academic years beginning autumn 2004 and 2005, and published in 2011.[9]

Suppose, in our trial, that researchers have an opportunity to visit each school in each of two distinct academic years, and that on each visit they assess 50 children from a single year-group (aged 9-10 years) in that school. In other words, there may be up to two, repeated, independent cross-sections from each school, over two time periods. Let us assume a continuous primary outcome measure analysed with a linear mixed regression model. The school setting naturally suggests a multi-level model with three levels: school, academic year within school, and pupil within academic year.

Let us further suppose that, while participating schools are all happy to implement the intervention, they would find it much harder (either practically or politically) to de-implement it. Then we must consider sequences with unidirectional cross-over, or no cross-over at all. We will allow incomplete designs, so there are seven possible sequences to which we can randomise clusters (see the first two columns of Table 1, where "X" in a sequence denotes no school visit in that academic year, 0 denotes a visit to collect data under the control condition, and 1 denotes a visit to collect data under the intervention condition).

How many schools should we randomise to each of these sequences in order to minimise the total number of visits to schools while achieving 90% power at the 5% significance level to detect a given treatment effect? Equivalently (and this is the problem we solve in this article): in what proportions should we allocate schools to these sequences in order to minimise the standard error of the resulting treatment effect estimate, if only a given number of school visits is possible?

**Table 1.** *Notation for the sequences in a two-period design, and representation of a design exhibiting centrosymmetry.*

| Sequences (X denotes no data collection) | | Number of clusters | BLUE coefficients | |
|---|---|---|---|---|
| Period 1 | Period 2 | | Period 1 | Period 2 |
| X | 1 | $n_{(0,\mathrm{X})}$ | | $y$ |
| 1 | X | $n_{(\mathrm{X},0)}$ | $u$ | |
| 1 | 1 | $n_{(0,0)}$ | $w$ | $z$ |
| 0 | 1 | $n_{(0,1)}$ | $-x$ | $x$ |
| 0 | 0 | $n_{(0,0)}$ | $-z$ | $-w$ |
| X | 0 | $n_{(\mathrm{X},0)}$ | | $-u$ |
| 0 | X | $n_{(0,\mathrm{X})}$ | $-y$ | |

## Optimal two-period designs

### ***Statistical model***

We write the sequence $\boldsymbol{s}$ as an ordered pair,

$$\boldsymbol{s} = (s_1, s_2) \epsilon \{(\mathrm{X}, 1), (1, \mathrm{X}), (1,1), (0,1), (0,0), (\mathrm{X}, 0), (0, \mathrm{X})\}.$$

where $s_t = 0{,}1$ or X, according as the cluster is observed and untreated (0), observed and treated (1) or unobserved (X) at time $t$. We assume the intervention will have the same effect, $\theta$, in every cluster in each period. This includes clusters allocated to sequence (1,1)—that is, we assume the effect of the intervention is maintained in period 2 if we introduce the intervention in period 1. This assumption of a maintained intervention effect is a standard starting point for modelling outcomes in stepped wedge trials.[10] We also allow for a change in mean outcome between period 1 and period 2 that is common to all clusters.

Suppose that $n_s$ clusters are allocated to sequence $\boldsymbol{s}$. Then the multi-level mixed model for the school breakfast example can be written as follows. If a visit takes place in period $t$ (= 1 or 2) of sequence $\boldsymbol{s}$, then $s_t = 0$ or 1, and there will be $m$ participants, $i = 1, \dots, m$, in each of $n_s$ clusters, $j = 1, \dots, n_s$, with outcomes

$$Y_{ijst} = \beta_t + \theta s_t + u_{js} + v_{jst} + e_{ijst}, \tag{1}$$

$u_{js}$ are independent identically distributed (IID) $\mathrm{N}(0, \sigma_u^2)$,

$v_{jst}$ are $\mathrm{IIDN}(0, \sigma_v^2)$,

$e_{ijst}$ are $\mathrm{IIDN}(0, \sigma_e^2)$,

$\beta_t$ is a fixed time effect.

The best linear unbiased estimate (BLUE) of the intervention effect, $\hat{\theta}$, will be a linear combination of the sequence-period means, which are defined for any sequence $\boldsymbol{s}$ with data-collection in period $t$ (i.e. for which $s_t = 0$ or 1) as

$$\bar{\bar{Y}}_{st} = \frac{1}{mn_s} \sum_{i,j} Y_{ijst}.$$

Each sequence-period mean $\bar{\bar{Y}}_{st}$ is itself a mean of $n_s$ cluster-period means, $\bar{Y}_{jst} = \frac{1}{m} \sum_i Y_{ijst}$. We want to minimise the variance of $\hat{\theta}$ by recruiting clusters to sequences subject to the constraint that the total number of cluster-periods with data collection is equal to a fixed number $N$:

$$n_{(\mathrm{X},1)} + n_{(1,\mathrm{X})} + 2n_{(1,1)} + 2n_{(0,1)} + 2n_{(0,0)} + n_{(\mathrm{X},0)} + n_{(0,\mathrm{X})} = N.$$

### *Centrosymmetry*

We will refer to a particular choice of the various $n_s$ as a design. We choose to work in a framework where the various $n_s$ can be in any ratios relative to one another. (If the $n_s$ are integers, this framework is the one we approach asymptotically as we increase $N$.)

The first thing to observe, in this framework, is that the optimal design, and the coefficients in the BLUE under this design, will be invariant under a transformation whereby we reverse the periods and swap the intervention and control condition (which we call a centrosymmetric transformation)[11]—see Table 1. For, suppose this were not the case. Then imagine a centrosymmetric transformation of the optimal design, and a centrosymmetric transformation of the coefficients of the BLUE under the optimal design. This must also constitute an optimal solution, since the optimisation problem looks the same under a centrosymmetric transformation. So now we have two minimal-variance solutions that are centrosymmetric opposites of each other. But then we can construct a minimal variance solution that is invariant under a centrosymmetric transformation by allocating half of our clusters to one of these designs, and half to the other, and averaging the two coefficients for each sequence in the combined design.

### *The two-period optimisation problem*

The centrosymmetric BLUE (see Table 1) may be written as

$$\hat{\theta} = y\bar{\bar{Y}}_{(X,1)2} + u\bar{\bar{Y}}_{(1,X)1} + w\bar{\bar{Y}}_{(1,1)1} + z\bar{\bar{Y}}_{(1,1)2} - x\bar{\bar{Y}}_{(0,1)1} + x\bar{\bar{Y}}_{(0,1)2} - z\bar{\bar{Y}}_{(0,0)1} - w\bar{\bar{Y}}_{(0,0)2} - u\bar{\bar{Y}}_{(X,0)2} - y\bar{\bar{Y}}_{(0,X)1}.$$

Let us denote the variance of each cluster-period mean, $\bar{Y}_{jst}$, as $\sigma^2$, and the correlation between two cluster-period means from the same cluster in different periods as $\rho$. It follows that the correlation between two sequence-period means from the same sequence in different periods is also $\rho$. This will turn out to be the parameter that determines the optimal design. In the multi-level mixed model (1), we can write $\rho$ in terms of the variance components:

$$\rho = \frac{\sigma_u^2}{\sigma_u^2 + \sigma_v^2 + \sigma_e^2/m}.$$

The parameter, $\rho$, can also be expressed in terms of the within-period intra-cluster correlation, $\rho_w$, and between-period intra-cluster correlation, $\rho_b$, which are sometimes adopted as an alternative parameterisation for the variance components in a stepped wedge model.[12,13]

Briefly:

$$\rho = \frac{m\rho_b}{1 + (m-1)\rho_w},$$

which may run the full gamut of values from zero to one, depending on whether $m\rho_b$ is small or large, and on the relative size of $\rho_b$ compared to $\rho_w$. (This parameterisation can also be helpful in guessing a plausible value for $\rho$ at the design stage.[8])

Then

$$\operatorname{var}\hat{\theta} = \sigma^2 \left[ \frac{1}{n_{(0,\mathrm{X})}}(2y^2) + \frac{1}{n_{(\mathrm{X},0)}}(2u^2) + \frac{1}{2n_{(0,0)}}\{4(z^2 + w^2 + 2\rho zw)\} + \frac{1}{n_{(0,1)}}\{2x^2(1-\rho)\} \right]. \quad (2)$$

This variance must be minimised subject to the following constraints.

To be unbiased, the expectation of $\hat{\theta}$ must equal $\theta$, and hence

$$x + y + z - w - u = 0;$$

$$x + y + z + w + u = 1.$$

Also, the total number of cluster-periods with data collection is fixed at $N$:

$$n_{(0,1)} + 2n_{(0,0)} + n_{(\mathrm{X},0)} + n_{(0,\mathrm{X})} = \frac{N}{2}. \quad (3)$$

The proof of the solution to this optimisation problem is given in the Appendix.

### *Two-period solution*

In summary, the solution has two distinct forms, depending on $\rho$:

1. For $\rho \leq \sqrt{3}/2$:

$$n_{(0,\mathrm{X})} = 0, \qquad n_{(\mathrm{X},0)} \propto \frac{1}{2}, \qquad n_{(0,0)} = 0, \qquad n_{(0,1)} \propto \frac{1}{2}(1-\rho)^{\frac{1}{2}}.$$

This design is illustrated in Figure 1(a) for $\rho \in \{0.1, 0.2, 0.3, 0.7, 0.8\}$. In this case

$$x = u = \frac{1}{2}, \qquad y = z = w = 0,$$

and the variance of the treatment effect estimator is

$$\operatorname{var}\hat{\theta} = \frac{\sigma^2}{N}\left(1 + (1-\rho)^{\frac{1}{2}}\right)^2.$$

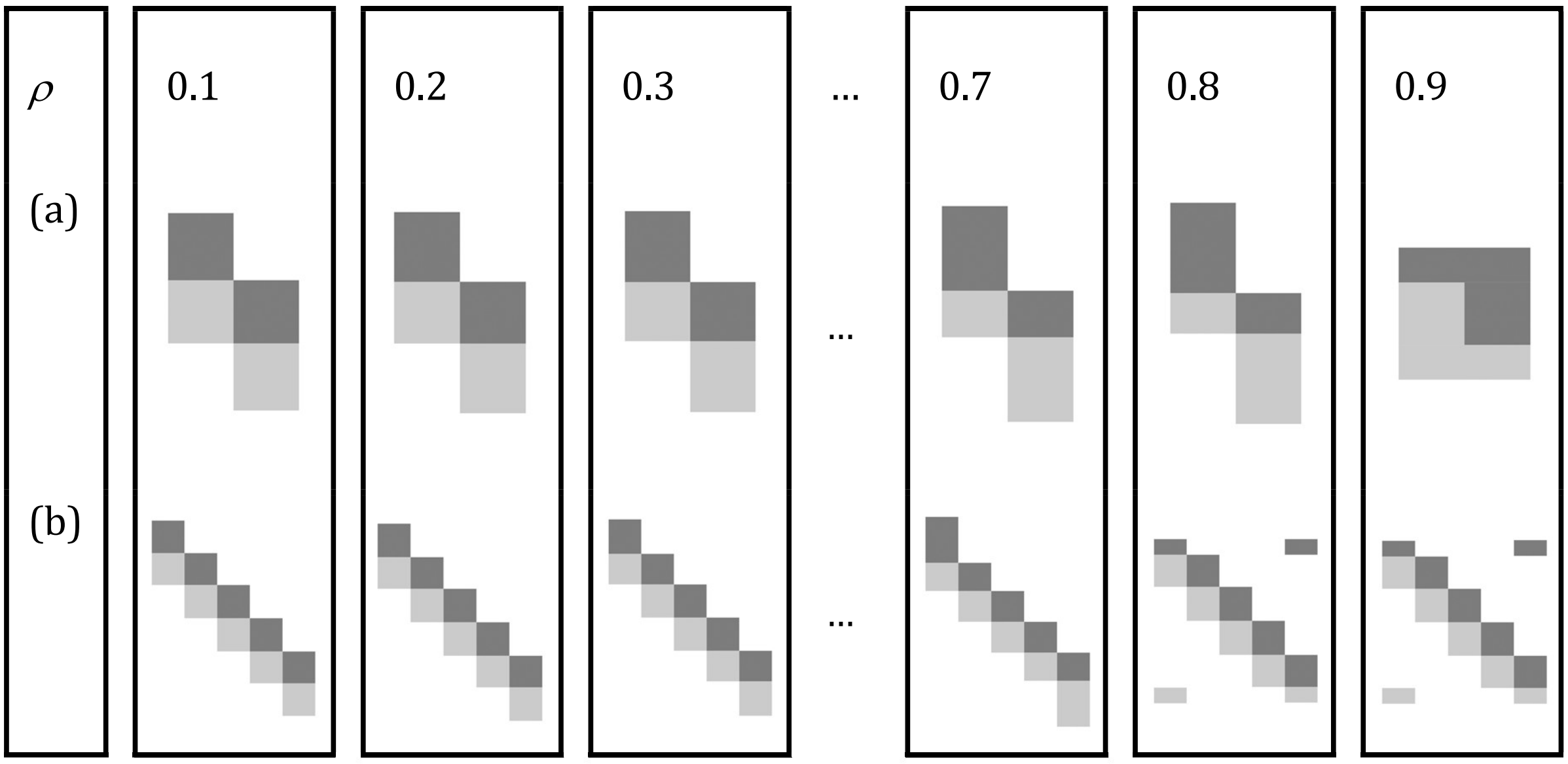

**Figure 1.** *(a) Schematic representation of the optimum two-period incomplete stepped wedge design, illustrated for different values of the correlation, $\rho$, between two cluster-period means from the same cluster in different periods. In each schematic, periods are shown as columns, and the height of each sequence (row) is in proportion to the weighting of allocations of clusters to that sequence. The total number of cluster-periods with data collection – or total shaded area – is the same for each schematic. At $\rho = \sqrt{3}/2 \approx 0.866$ the form of the optimal two-period design "flips" between the two distinct forms. (b) The conjectured solution to the optimisation problem for five periods, from which the form of the conjectured solution for multiple periods can be generalised.*

2. For $\rho \geq \sqrt{3}/2$:

$$n_{(0,\mathrm{X})} = 0, \qquad n_{(\mathrm{X},0)} = 0, \qquad n_{(0,0)} \propto \frac{1}{2}(1-\rho)^{\frac{1}{2}}, \qquad n_{(0,1)} \propto \rho(1-\rho)^{\frac{1}{2}}.$$

This design is illustrated in Figure 1(a) for $\rho = 0.9$. In this case

$$x = \rho, \qquad u = 0, \qquad y = 0, \qquad z = \frac{1}{2} - \rho, \qquad w = \frac{1}{2},$$

and the variance of the treatment effect estimator is

$$\operatorname{var}\hat{\theta} = \frac{4\sigma^2}{N}(1+\rho)^2(1-\rho).$$

Figure 2(a) plots these variances against $\rho$, illustrating the superiority of the design in case (1) for smaller $\rho$, and the superiority of the design in case (2) for larger $\rho$. The two solutions are equally good when

$$1 + (1-\rho)^{\frac{1}{2}} = 2(1+\rho)(1-\rho)^{\frac{1}{2}}.$$

This happens for $\rho = \sqrt{3}/2 \approx 0.8660$.

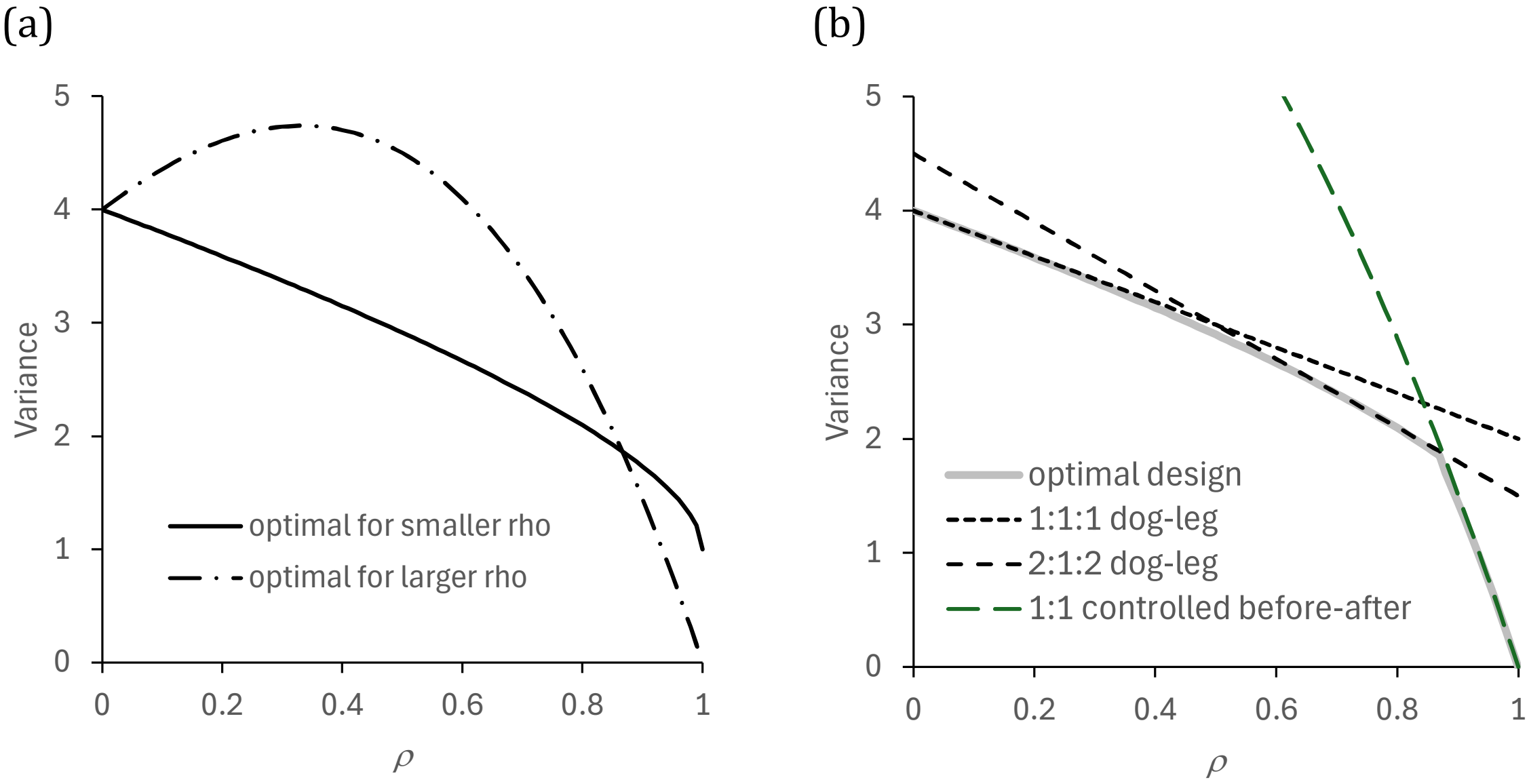

**Figure 2.** *For the two-period design problem, the variance of the best linear unbiased treatment effect estimator (i.e. the minimum variance) plotted against the correlation, $\rho$, between the two cluster-period means from the same cluster in different periods. Variances are proportional to $\sigma^2/N$, where $\sigma^2$ is the variance of each cluster-period mean and $N$ is the total number of cluster-periods. Here, the variance axis is scaled so that $\sigma^2/N = 1$. The minimum variance is plotted for particular choices of two-period design: (a) the optimal design for $\rho \leq \sqrt{3}/2$ (case (1) in the text) and the optimal design for $\rho \geq \sqrt{3}/2$ (case (2) in the text); (b) a dog-leg design with allocation ratios 1:1:1, a dog-leg design with allocation ratios 2:1:2, and a controlled before-after design with allocation ratio 1:1, with the minimum variance under the optimal design shown for comparison.*

### *Conjectured solution for more than two periods*

We have not found a proof of the solution to the analogous optimisation problem with more than two periods, but we have explored potential solutions using a "greedy search" algorithm to explore the space of possible designs.[14] Briefly, the user specifies an admissible starting design, and the algorithm considers incremental changes in which a small proportion of clusters from one sequence is swapped to another, but preserving the total number of cluster-periods where there is data collection. The algorithm chooses the swap that maximises the increase in precision of the treatment effect estimate, calculated using generalised least squares theory and numerical matrix inversion to obtain the BLUE, and continues iteratively until no further improvement is possible. We assumed a block-exchangeable structure for the intra-cluster correlation,[15] *i.e.* that the correlation between two cluster-period means from the same cluster in two different periods is a constant, $\rho$. Code written in Mata (a C-style, matrix programming language within Stata (Stata Corp, College Station, Texas USA)) is available online (https://github.com/richard-hooper/optimal-incomplete-SW). As this is not an exhaustive search of the design space it is not guaranteed to find a global optimum, though there is some reassurance to be had from starting the algorithm off in different parts of the design space and seeing it identify the same solution.

An exploration using our greedy search algorithm of solutions for three, four or five periods at particular values of $\rho$, and with two alternative starting designs, led us to conjecture a multi-period solution that generalises the form of the optimal design for two periods. We emphasise that the use of the search algorithm does not constitute a proof: our solution is presented as an unproven conjecture. We illustrate the designs indicated by our algorithm in the five-period case in Figure 1(b). We conjecture that in the multi-period optimisation problem, the optimum for $\rho$ below some cut-off is a design restricted to the sequences $(1, \mathrm{X}, \mathrm{X}, \dots, \mathrm{X})$, $(\mathrm{X}, \mathrm{X}, \dots, \mathrm{X}, 0)$, and to the sequences $(0, 1, \mathrm{X}, \mathrm{X}, \dots, \mathrm{X})$, $(\mathrm{X}, 0, 1, \mathrm{X}, \mathrm{X}, \dots, \mathrm{X})$, $(\mathrm{X}, \mathrm{X}, 0, 1, \mathrm{X}, \mathrm{X}, \dots, \mathrm{X})$, …, $(\mathrm{X}, \mathrm{X}, \dots, \mathrm{X}, 0, 1)$, and for $\rho$ above the cut-off is a design restricted to the sequences $(1, \mathrm{X}, \mathrm{X}, \dots, \mathrm{X}, 1)$, $(0, \mathrm{X}, \mathrm{X}, \dots, \mathrm{X}, 0)$, and to the sequences $(0, 1, \mathrm{X}, \mathrm{X}, \dots, \mathrm{X})$, $(\mathrm{X}, 0, 1, \mathrm{X}, \mathrm{X}, \dots, \mathrm{X})$, $(\mathrm{X}, \mathrm{X}, 0, 1, \mathrm{X}, \mathrm{X}, \dots, \mathrm{X})$, …, $(\mathrm{X}, \mathrm{X}, \dots, \mathrm{X}, 0, 1)$.

# Discussion

## *Optimal two-period designs*

The optimal two-period design for $\rho < \frac{\sqrt{3}}{2}$, as illustrated in Figure 1(a), has a form described elsewhere as a "dog-leg", referring to its crooked shape.[16] The treatment effect estimator in this case is simply the average of the two within-period contrasts of intervention and control:

$$\hat{\theta} = \frac{1}{2}\left[\left(\bar{\bar{Y}}_{(1,\mathrm{X})1} - \bar{\bar{Y}}_{(0,1)1}\right) + \left(\bar{\bar{Y}}_{(0,1)2} - \bar{\bar{Y}}_{(\mathrm{X},0)2}\right)\right].$$

This can be rewritten as half of the difference between two contrasts of period 2 with period 1:

$$\hat{\theta} = \frac{1}{2}\left[\left(\bar{\bar{Y}}_{(0,1)2} - \bar{\bar{Y}}_{(0,1)1}\right) - \left(\bar{\bar{Y}}_{(X,0)2} - \bar{\bar{Y}}_{(1,X)1}\right)\right].$$

The first term in this difference is a paired contrast *within* clusters in sequence $(0, 1)$; the second term is un unpaired contrast *between* clusters in sequences $(X, 0)$ and $(1, X)$. Some efficiency is gained from the correlation between the sequence-period means in the paired contrast. Counter-intuitively, though, as this correlation increases, the most efficient way to allocate clusters to sequences is to put more and more data collection effort into the contrast that is harder to estimate accurately—the unpaired contrast.

Eventually as $\rho$ increases a critical value is reached, beyond which the form of the optimal design "flips" dramatically. Beyond this point it becomes more efficient to concentrate on within-cluster comparisons of all clusters, at the cost of collecting data in both time periods in all clusters. The optimal design for $\rho > \frac{\sqrt{3}}{2}$ is essentially a controlled before-after design, $(0, 1), (0, 0)$, sitting underneath its centrosymmetric transformation, $(1, 1), (0, 1)$. In a controlled before-after design, for $\rho$ close to 1, the optimal treatment effect estimator is close to being a difference-in-differences estimator—that is, an estimator that subtracts the difference between period 2 and period 1 in sequence $(0, 0)$ from the difference between period 2 and period 1 in sequence $(0, 1)$.[17] This difference-in-differences approach to estimating the treatment effect, with its *positive* coefficient for $\bar{\bar{Y}}_{(0,0)1}$ (data collected under the control condition), is fundamentally different to the approach of the dog-leg, which makes use of within-period contrasts of intervention and control.

This duality of estimation approaches is related to the idea of "horizontal" and "vertical" comparisons in complete stepped wedge designs, where one can always decompose the treatment effect estimator into a so-called "horizontal" comparison (the optimal estimator in the presence of fixed cluster effects) and a "vertical" comparison made up of contrasts between intervention and control in each period.[18] This theory breaks down in the case of incomplete designs, where the treatment effect may not be identifiable with a fixed effects model (as is the case for the dog-leg design, for example).

### *A simple design choice*

We have allowed designs in which clusters could be allocated to sequences in any ratios, but even allowing only a very restricted choice of designs we can achieve close-to-optimal performance over the range of correlations $0 < \rho < 1$. Figure 2(b) shows the variance of the treatment effect estimator under three, simple, two-period designs: a dog-leg with allocation ratios 1:1:1, a dog-leg with allocation ratios 2:1:2, and a controlled before-after design with allocation ratio 1:1. For comparison, Figure 2(b) also shows the variance achieved under the optimal two-period design, as shown in Figure 2(a). Figure 2(b) shows that we could select a 1:1:1 dog-leg when $\rho < 0.5$, a 2:1:2 dog-leg when $0.5 < \rho < 0.875$, or a 1:1 controlled before-after design when $\rho > 0.875$, and never stray far from optimal performance (the cut-offs for $\rho$ can be determined from formulae for the respective variances under different designs). This offers a simple and practical approach to the design of two-period trials.

The simplicity of this design strategy also illustrates that there is generally no risk of sacrificing efficiency through a poor choice of design based on an inaccurate guess at $\rho$, unless we think $\rho$ may be close to $\sqrt{3}/2$, in which case we might end up making an inefficient choice between a dog-leg and a controlled before-after design. To quantify further: if we believed $\rho$ to be $\sqrt{3}/2$ (at which value the two design forms in Figure 2(a) are jointly optimal) but if we had over-estimated or under-estimated this value by just 0.02, then using the suboptimal form in place of the optimal form would inflate the variance by 8% (if $\rho$ is over-estimated) or by 10% (if $\rho$ is under-estimated).

### *Motivating example*

A detailed discussion of sample size in our motivating example is available in an earlier reference.[8] Note that sample size calculations of this kind typically involve making informed guesses about variance components or intra-cluster correlation coefficients, and applying sample size formulae based on asymptotic theory. Using figures from the original sample size calculation reported for the Welsh Primary School Free Breakfast Initiative ,[9] and making further assumptions, these authors posited a value of 0.4040 for the correlation, $\rho$, between cluster-period means from the same cluster. They calculated the sample size needed to detect an effect size (ratio of mean difference to standard deviation) of 0.11 with 80% power at the 5% significance level, under a variety of alternative designs. In particular, with a 1:1:1 dog-leg design (which we have shown in this article to be close to optimal in terms of minimising the total number of cluster-periods with data collection), the required sample size was 63 schools, resulting in 84 school visits in total (4,200 children sampled – 50 at each school visit). In comparison, a 1:1 controlled before-after design would require 88 schools and 176 school visits in total (8,800 children).

### *Further reflections on optimal incomplete stepped wedge trial designs*

Our conjectured solution to the optimisation problem for more than two periods suggests one way that the two-period solution, which flips from a dog-leg to a complete two-period design, might be generalised. In this conjectured multi-period solution, the optimal design flips from a zig-zagging diagonal pattern, sometimes referred to in the stepped wedge literature as a "staircase" design,[19] to something similar but with the addition of data collection in the cluster-periods in the off-diagonal corners. These off-diagonal corners have been named the "hot-spots" of the diagram, since (along with the leading diagonal) they are the regions that contribute the most information in a complete design.[20] While other authors have focused on the advantages of particular kinds of incomplete design, such as a "basic staircase" (one control period followed by one intervention period in each sequence, sequences equally weighted),[19] our investigation has been into the more general problem of choosing an optimal weighting for any or all of the possible complete or incomplete sequences.

In the multi-period case we assumed a categorical effect of period and a block exchangeable intra-cluster correlation. With a relatively small number of discrete periods we suspect that most analysts would be content to model the period effect as categorical. Even if the true relationship were simpler (linear in period, say) then this analysis model would still be correctly specified and our conclusions would still hold. As discussed for the two-period solution, the advantage of a "flipped" design with hot-spots rests on trading off the benefit of a within-cluster comparison between the (correlated) first and last period against the cost of collecting data in both of these periods. A conventional staircase design, meanwhile, draws its power from the correlation between outcomes in consecutive periods. If we have mis-specified the form of the intra-cluster correlation – if, in fact, it decays with increasing separation between periods—then the threshold for the correlation, $\rho$, above which we would consider including hot-spots would be raised. We do not have a formal expression for this conjectured threshold, so this discussion must remain qualitative for now. Future numerical investigations could address this issue in more detail. Investigations of optimal incomplete designs where the time effect is modelled as a polynomial have thrown a spotlight on a variety of unusual designs, including designs with hot-spots, and multi-phase before-after deigns.[22]

In practice, there is always a danger that clusters will be more likely to drop out of some sequences than others following randomisation. Not only would this impair the efficiency of the design, but it would also introduce a risk of bias. The CONSORT reporting guideline extension to stepped wedge trials recommends that authors report, for each sequence, details of losses and exclusions of both clusters and participants, with reasons.

We have targeted the design that minimises the total number of cluster-periods with data collection for given precision. This design also minimises the total number of individual participants, on the assumption that we either collect data on $m$ participants or on none in each cluster-period. What if we could vary the number of participants between 0 and $m$ in each cluster-period: what design would minimise the total number of participants for given precision? This question is currently unexplored, but may not always have sensible-looking solutions. Observe how our dog-leg design could be made more efficient by having smaller clusters, and more of them, in sequences $(1, \mathrm{X})$ and $(\mathrm{X}, 0)$, reducing the variance-inflating effects of clustering on the unpaired contrast between time periods. This leads ultimately to a design with a very large number of "clusters" each of one participant—hundreds of schools where we sample just one pupil, for example. A real-life stepped wedge design problem will involve trading off the costs of various design elements—cost per cluster, cost per period, cost per cluster-period of data collection, cost per individual participant, *etc.*—subject also to upper limits on the number of clusters, number of periods, number of participants per cluster-period, *etc*. Nevertheless, our solutions add significantly to our understanding of optimal stepped wedge design, and make an original contribution to the study of the design of experiments.

## Authors' contributions

RH conceived the article and conducted the numerical investigations that suggested the solution. AG proved the result for two-period designs. RH wrote the first draft of the manuscript and prepared the Figures. Both authors contributed to the final version of the manuscript. There was no external funding for the work.

## Appendix: Proof of the two-period solution

First, we make use of the general result that, for any numbers $b_1, b_2, \dots, b_K$, the minimum value of

$$\sum_{i=1}^{K} \frac{b_i^2}{m_i}$$

over $m_i \geq 0$, subject to the constraint $\sum m_i = M$, is

$$\frac{1}{M}\left(\sum_{i=1}^{K} |b_i|\right)^2,$$

and this happens when

$$m_i = M \frac{|b_i|}{\Sigma_j |b_j|} \propto |b_i|, \qquad i = 1, \dots, K.$$

We want to apply this result to find the values of the $n_s$ that minimise the variance of the treatment effect estimator set out in equation (2) of the main article, subject to the constraint (3). We reproduce (2) and (3) below:

$$\operatorname{var} \hat{\theta} = \sigma^2 \left[\frac{1}{n_{(0,\mathrm{X})}}(2y^2) + \frac{1}{n_{(\mathrm{X},0)}}(2u^2) + \frac{1}{2n_{(0,0)}}\{4(z^2 + w^2 + 2\rho zw)\} + \frac{1}{n_{(0,1)}}\{2x^2(1-\rho)\}\right]; \qquad (2)$$

$$n_{(0,1)} + 2n_{(0,0)} + n_{(\mathrm{X},0)} + n_{(0,\mathrm{X})} = \frac{N}{2}. \qquad (3)$$

This leads to

$$n_{(0,X)} \propto |y|, n_{(X,0)} \propto |u|, 2n_{(0,0)} \propto \sqrt{2}\sqrt{z^2 + w^2 + 2\rho zw}, n_{(0,1)} \propto |x|(1-\rho)^{\frac{1}{2}},$$

whereupon

$$\operatorname{var}\hat{\theta} = \frac{4\sigma^2}{N}\left[|y| + |u| + \sqrt{2}\sqrt{z^2 + w^2 + 2\rho zw} + |x|(1-\rho)^{\frac{1}{2}}\right]^2.$$

We must minimise the expression in square brackets

$$F(x,y,z,w,u) = |y| + |u| + \sqrt{2}\sqrt{z^2 + w^2 + 2\rho zw} + |x|(1-\rho)^{\frac{1}{2}}$$

subject to the constraints on $x, y, z, w$ and $u$ which arise from the unbiasedness of the treatment effect estimator (see main article). These constraints reduce to

$$x + y + z = w + u = \frac{1}{2}.$$

First, note that whenever $y \neq 0$,

$$F(x+y, 0, z, w, u) < F(x, y, z, w, u).$$

This shows that $y = 0$ at the minimum, so that $n_{(0,X)} = 0$ for any value of $\rho$.

Now use the constraints on $x, y, z, w, u$ to express $F$ in terms of $x$ and $u$:

$$F(x,u) = \sqrt{2}\sqrt{\left(x-\frac{1}{2}\right)^2 + \left(u-\frac{1}{2}\right)^2 + 2\rho\left(x-\frac{1}{2}\right)\left(u-\frac{1}{2}\right)} + |u| + |x|(1-\rho)^{\frac{1}{2}}.$$

To minimise this, consider the behaviour of $F$ along a ray through the point $\left(\frac{1}{2},\frac{1}{2}\right)$—i.e. set

$$\left(u-\frac{1}{2}\right) = \lambda\left(x-\frac{1}{2}\right), \text{ for some } \lambda, \ -\infty < \lambda < \infty.$$

(We also need to check the vertical ray $x = \frac{1}{2}$, corresponding to $\lambda = \infty$.)

Along such a ray we have

$$F(x,u) = F_\lambda(x) = \sqrt{2}\left|x - \frac{1}{2}\right|\sqrt{1+\lambda^2+2\rho\lambda} + \left|\frac{1}{2} + \lambda\left(x - \frac{1}{2}\right)\right| + |x|(1-\rho)^{\frac{1}{2}}.$$

Clearly $F_\lambda(x)$ is piece-wise linear in $x$, and goes to $\infty$ as $x \to \pm\infty$. Therefore, its minimum must occur at one of its knots—i.e. $x = 0$, ½ or ½$(1 - \lambda^{-1})$. The last value corresponds to $u = 0$.

It follows that, at the minimum value of $F(x,u)$, we must have at least one of

$$x = 0, u = 0, x = \frac{1}{2}, \text{ or } u = \frac{1}{2}.$$

The corresponding values of $F$ are:

(i) $x = 0,\ F(0,u) = \sqrt{2}\sqrt{\left(u - \frac{1}{2(1+\rho)}\right)^2 + \frac{1}{4(1-\rho^2)}} + |u|$

(ii) $u = 0,\ F(x,0) = \sqrt{2}\sqrt{\left(x - \frac{1}{2(1+\rho)}\right)^2 + \frac{1}{4(1-\rho^2)}} + |x|(1-\rho)^{\frac{1}{2}}$

(iii) $x = \frac{1}{2},\ F\left(\frac{1}{2},u\right) = \sqrt{2}\left|u - \frac{1}{2}\right| + |u| + \frac{1}{2}(1-\rho)^{\frac{1}{2}}$

(iv) $u = \frac{1}{2},\ F\left(x,\frac{1}{2}\right) = \sqrt{2}\left|x - \frac{1}{2}\right| + \frac{1}{2} + |x|(1-\rho)^{\frac{1}{2}}$

Note first that $F(0,u) > F(u,0)$ unless $u = 0$. Hence, case (i) is superseded by case (ii).

In case (ii), the minimum must occur for a positive value of $x$, because $F(x,0)$ is decreasing in $x$ when $x < 0$. For $x > 0$ the function is differentiable with a turning value at $x = \rho$. This corresponds to a minimum at which $x = \rho, u = 0$ with

$$\left(\frac{N}{4\sigma^2}\right)\text{var}\,\hat{\theta} = F(\rho,0) = (1+\rho)(1-\rho)^{\frac{1}{2}}$$

.

The expressions (iii) and (iv), are both minimised at $x = u = \frac{1}{2}$, whereupon

$$\left(\frac{N}{4\sigma^2}\right)\text{var}\,\hat{\theta} = F\left(\frac{1}{2},\frac{1}{2}\right) = \frac{1}{2}\left[1 + (1-\rho)^{\frac{1}{2}}\right].$$

In summary, the global minimum occurs at $y = 0, x = u =$ ½ provided that

$$1 + (1-\rho)^{\frac{1}{2}} \leq 2(1+\rho)(1-\rho)^{\frac{1}{2}}. \quad \text{(A1)}$$

Otherwise, when the inequality is reversed, the minimum occurs at $y = u = 0, x = \rho$. The inequality (A1) is satisfied whenever $\rho \leq \sqrt{3}/2$, as may be shown by setting $q = (1-\rho)^{\frac{1}{2}}$ and factorising the resulting cubic.